\documentclass[11pt,letterpaper]{article}

\usepackage{typearea}
\typearea{15}

\usepackage{theorem,latexsym,graphicx}
\usepackage{amsmath,amssymb,enumerate}
\usepackage{xspace}
\usepackage{bm}
\usepackage{ifpdf}
\usepackage{shadow,shadethm,color}
\usepackage[compact]{titlesec}
\usepackage{algorithm}
\usepackage[noend]{algorithmic}
\usepackage{subfigure}
\usepackage{verbatim}

\definecolor{Darkblue}{rgb}{0,0,0.4}
\definecolor{Brown}{cmyk}{0,0.81,1.,0.60}
\definecolor{Purple}{cmyk}{0.45,0.86,0,0}
\newcommand{\mydriver}{hypertex}
\ifpdf
 \renewcommand{\mydriver}{pdftex}
\fi
\usepackage[breaklinks,\mydriver]{hyperref}
\hypersetup{colorlinks=true,
            citebordercolor={.6 .6 .6},linkbordercolor={.6 .6 .6},%
citecolor=Darkblue,urlcolor=black,linkcolor=Darkblue,pagecolor=black}
\newcommand{\lref}[2][]{\hyperref[#2]{#1~\ref*{#2}}}


\newtheorem{theorem}{Theorem}[section]
\newtheorem{definition}[theorem]{Definition}

\newtheorem{lemma}[theorem]{Lemma}
\newshadetheorem{lemmashaded}[theorem]{Lemma}

\newtheorem{claim}[theorem]{Claim}

\newtheorem{corollary}[theorem]{Corollary}
\newenvironment{proof}{

\noindent{\bf Proof:}}
{\hfill$\blacksquare$

}

\newenvironment{Myquote}{\par\begingroup
\addtolength{\leftskip}{1em} \rightskip\leftskip }{\par
\endgroup
}

\newcommand{\junk}[1]{}
\newcommand{\ignore}[1]{}

\newcommand{\opt}{\ensuremath{{\sf OPT}}\xspace}

\newcommand{\mst}{\ensuremath{{\sf MST}}\xspace}
\newcommand{\htsp}{\ensuremath{{\sf HetTSP}}\xspace}
\newcommand{\hvrp}{\ensuremath{{\sf HetCVRP}}\xspace}
\newcommand{\cvrp}{\ensuremath{{\sf CVRP}}\xspace}
\newcommand{\tsp}{\ensuremath{{\sf TSP}}\xspace}
\newcommand{\p}{\ensuremath{\mathcal{P}}\xspace}

\newcommand{\levelPrim}{{\sf Level-Prim}\xspace}
\newcommand{\last}{{\sf LAST}\xspace}

\def\T{\mathcal{T}}
\def\subt{\mathcal{S}}
\def\dangle{\sf{Dangle}}
\newcommand{\sse}{\subseteq}

\newcommand{\initOneLiners}{%
    \setlength{\itemsep}{0pt}
    \setlength{\parsep }{0pt}
    \setlength{\topsep }{0pt}
}
\newenvironment{OneLiners}[1][\ensuremath{\bullet}]
    {\begin{list}
        {#1}
        {\initOneLiners}}
    {\end{list}}

\newcounter{note}[section]

\title{Capacitated Vehicle Routing with Non-Uniform Speeds}
\author{Inge Li G{\o}rtz\thanks{Technical University of Denmark.} \and Marco Molinaro\thanks{Tepper School of Business, Carnegie Mellon University. Supported in part by NSF grant CCF-0728841.}
\and Viswanath Nagarajan\thanks{IBM T.J. Watson Research Center.} \and R. Ravi$^\dagger$}
\date{}

\begin{document}
\begin{titlepage}

\maketitle

\begin{abstract}
The {\em capacitated vehicle routing problem} (CVRP)~\cite{vrp-book} involves distributing (identical) items from a depot to a
set of demand locations, using a single capacitated vehicle. We study a generalization of
this problem to the setting of multiple vehicles having non-uniform speeds (that we call {\em Heterogenous CVRP}), and present a constant-factor approximation
algorithm.

The technical heart of our result lies in achieving a constant approximation to the following TSP variant (called {\em
Heterogenous TSP}). Given a metric denoting distances between vertices, a depot $r$ containing $k$ vehicles having respective
speeds $\{\lambda_i\}_{i=1}^k$, the goal is to find a tour for each vehicle (starting and ending at $r$), so that every
vertex is covered in some tour and the maximum completion time is minimized. This problem is precisely Heterogenous
CVRP when vehicles are uncapacitated.

The presence of non-uniform speeds introduces difficulties for employing standard tour-splitting techniques. In order to get a better understanding of this technique in our context, we appeal to ideas from the 2-approximation for scheduling in parallel machine of Lenstra et al.~\cite{lenstraST}. This motivates the introduction of a new approximate MST construction called {\em Level-Prim}, which is related to {\em Light Approximate Shortest-path Trees}~\cite{KRY95}. The last component of our algorithm involves partitioning the Level-Prim tree and matching the resulting parts to vehicles. This decomposition is more subtle than usual since now we need to enforce correlation between the size of the parts and their distances to the depot.


\end{abstract}

\thispagestyle{empty}

\end{titlepage}

\section{Introduction}
The capacitated vehicle routing problem (\cvrp) is an extensively studied combinatorial optimization problem (see e.g.,
the book~\cite{vrp-book} and references therein). \cvrp is defined on a metric space $(V,d)$, where $V$ is a finite set of
locations/vertices and $d:V\times V\rightarrow \mathbb{R}_+$ a distance function that is symmetric and satisfies
triangle inequality. There is a depot  vertex $r\in V$ that contains an infinite supply of an identical item, and each
vertex $u\in V$ demands some units $q_u$ of this item. A single vehicle of capacity $Q\ge 0$
is used to distribute the items. The objective is to find a minimum length tour of the vehicle that satisfies all
demands subject to the constraint that the vehicle carries at most $Q$ units at any time.

\cvrp is closely related to the Traveling Salesman Problem (\tsp). It is clear that \cvrp reduces to \tsp in the absence of
capacity constraint. More interestingly, a reverse relation is also known---essentially the best known approximation algorithm for \cvrp~\cite{haimovichRK}
achieves a guarantee of $\rho+1$, where $\rho$ is the best approximation ratio for \tsp.

In practice, it is natural to have a fleet of \emph{multiple} vehicles that can run in parallel. The objective can then be to either minimize the sum of completion times of all the vehicles or to minimize the maximum completion time over all vehicles (or the makespan of the routing). Furthermore the vehicles can all be identical (same speed) or heterogeneous (have different speeds). In either case, it is not hard to see that the total completion time objective reduces to the usual \cvrp on a single maximum-speed vehicle, and constant-factor approximation algorithms readily follow.

When the objective is to minimize the makespan with identical vehicles, ideas for approximating the regular \cvrp problem using a tour-splitting heuristic introduced by Frederickson et al.~\cite{frederickson} can be easily adapted to derive a constant-factor approximation algorithm (see below).

This motivates the {\em Heterogenous Capacitated Vehicle Routing Problem} (\hvrp) that we consider. In this problem, a fleet of $k$ vehicles with {\em non-uniform speeds} and uniform capacities is initially located at the depot vertex $r$.
The objective is to satisfy the demands subject to the capacity constraints while minimizing the makespan.
Our main result is a constant-factor approximation algorithm for \hvrp.

Most of our algorithmic ideas lie in solving the special case of \hvrp when there is no capacity constraint.
This problem, which we call \htsp, is a generalization of \tsp that might be of independent interest. For most of
this paper, we will focus on obtaining a constant-factor approximation for \htsp.

\subsection{Previous Techniques} \label{sec:previousTech}
{\bf Tour-splitting solutions:} To illustrate the use of known techniques, we outline how to obtain a constant-factor approximation algorithm for \htsp with uniform
speeds~\cite{frederickson}.
First, notice that the union of the tours of $\opt$ connects all vertices, and hence a minimum spanning tree has length at most $k\cdot \opt$. Then consider an \mst, duplicate its edge and take an Euler tour $C$, which is of length
$d(C) \le 2k\cdot \opt$. Now split $C$ into $k$ segments of lengths at most $\frac{d(C)}{k}$ by removing edges. Finally, the
tour for the $i^{th}$ vehicle is obtained by connecting both endpoints of the $i^{th}$ segment of $C$ to the depot.
Since twice the distance from the depot to any vertex is a lower bound on $\opt$, the length of each tour is at most $3 \cdot \opt$ and hence this solution is a 3-approximation. We remark that this can be extended to obtain an $O(1)$-approximation for \hvrp with uniform speeds (e.g., using Theorem~\ref{th:vrptotsp} in Section~\ref{sec:hvrp}).

At a very high level, this strategy has two main components: (1) Partitioning an \mst into manageable-sized connected
parts; (2) assigning these parts to vehicles. This simple idea---which was already present in the 70's---is the central
piece of many heuristics and approximations for
vehicle routing problems (e.g.,~\cite{frederickson,haimovichRK, Even2004, Arkin20061, GNR,ag87,ag90}).
However, it is not clear how to employ this technique in the presence of vehicles with multiple speeds. This is because the two
main components now need some correlation: a small part of the \mst, which should be assigned to a slower vehicle, must also be relatively closer to the depot in order to be reachable by this vehicle.

{\bf Set-cover based solutions:} For \htsp with non-uniform speeds, previous approaches seem to give only a logarithmic approximation, as follows.
Guess the optimal makespan \opt (within a constant factor). If each vehicle of speed $s$ is given a length budget of
$s\cdot \opt$, then the vehicles can collectively cover all vertices. Using an approximation algorithm for
$k$-\mst~\cite{garg05} (or the related orienteering problem~\cite{blum07,chekuriKP08}) within a maximum-coverage framework (see e.g..~\cite{chekuriK04}), we
can obtain tours of length \opt that cover a constant fraction of all vertices. Repeating this
coverage step until all vertices are covered gives a solution to \htsp of makespan $O(\log n)\cdot\opt$. The intrinsic problem of this approach is that it is too general---in fact, the above algorithm also yields a logarithmic approximation even in the setting where the metric faced by each vehicle is arbitrary (instead of just scaling by its speed), and this generalization of
\htsp can be shown to be set-cover hard. It is unclear whether the approximation of this set-covering based approach can be improved for \htsp.

\subsection{Results, Techniques and Outline}

We extend the first tour-splitting approach described above to obtain the following result.

\begin{theorem}
There are constant-factor approximation algorithms for \htsp and \hvrp.
\end{theorem}

In order to obtain the approximation for \htsp, we abstract the requirements of the two components in the tour-splitting strategy. As a preprocessing step, we round the speeds of vehicles to powers of two and guess the optimum makespan $M$. First, we specify conditions which guarantee that a collection of $r$-rooted trees is ``assignable'', that is, each vehicle can visit the nodes of the trees assigned to it within time $O(M)$ (Definition~\ref{def:assignable}). 
The conditions in Definition~\ref{def:assignable} are based on the LP to obtain a 
2-approximation for {\em scheduling in unrelated parallel machines} by Lenstra et al.~\cite{lenstraST}.


Secondly, instead of partitioning an \mst as in the previous section, we consider more structured spanning trees which we call \levelPrim trees.
%
%
%
Consider grouping the vertices ordered according to their distance from $r$ into 
levels, where the $i$th level includes all vertices within distance $2^i M$.\footnote{Notice that given the rounding of vehicle speeds to powers of two, vertices in level $i$ can only be served by vehicles of speed $2^i$ or higher given the makespan bound $M$.} 
The \levelPrim tree is simply the tree resulting from running Prim's algorithm with the restriction that all nodes in a level are spanned before starting to pull in nodes from the next.


A \levelPrim tree has two important properties: (i) The vertices along every root-leaf path are monotonically nondecreasing in level
and (ii) For every suffix of levels,
the subgraph induced on it costs at most $O(1)$ times its induced \mst. The first condition, which is the departing point from {\mst}s, greatly simplifies the decomposition procedure carried in the next step. The second property is related to the assignability conditions in Definition~\ref{def:assignable} and guarantees that the we can decompose a \levelPrim tree into an assignable collection. These properties are formalized in Theorem~\ref{thm:lPrim}.

The \levelPrim construction combine both \mst and shortest-path distances from a root, so it is not surprising that this structure is related to {\em Light Approximate Shortest-Path Trees} (\last) introduced by Khuller et al.~\cite{KRY95}. Indeed, we use the existence of
a suitably defined \last in proving Theorem~\ref{thm:lPrim}. We remark, however, that the properties guaranteed by {\last}s are not enough for our purposes (see Section \ref{sec:levelPrim}).

The third main component of our approximation for \htsp is decomposing \levelPrim into an assignable collection of $r$-rooted trees. Roughly, we partition the edges of \levelPrim into subtrees while ensuring that each subtree consisting of vertices in levels up to $i$ (and hence is at a distance of about $2^i M$ from the root) also has length approximately $2^i M$, and thus can be assigned to a vehicle of speed about $2^i$. This partition, which relies on the two properties of \levelPrim, gives a collection of \emph{unrooted} trees which is assignable.
Due to the length of these trees, the extra distance to connect them to the root $r$ can be charged to their edges, hence this collection can be turned into a $r$-rooted assignable collection.


In order to obtain an approximation to \hvrp, we reduce this problem to approximating \htsp in a suitably modified metric space. This new distance function encodes any additional trips to and from the root that a vehicle has to make if it runs out of capacity. The exact transformation is presented in Section~\ref{sec:hvrp}.

\subsection{Related Work}
For the \cvrp, the best known approximation ratio~\cite{haimovichRK} is essentially $\rho+1$ where $\rho$ is
the best guarantee for \tsp.
The current best values for $\rho$ are $\rho=\frac32$ for general metrics~\cite{christofides}, and $\rho=1+\epsilon$ (for any
constant $\epsilon>0$) for constant dimensional Euclidean metrics~\cite{arora,mitchell}. This has been improved slightly to
$1+\rho\cdot (1-\frac1Q)-\frac{1}{3Q^3}$ when $Q\ge 3$~\cite{bdo07}. Recently, Das and Mathieu~\cite{DasM10} gave a
quasi-polynomial time approximation scheme
for \cvrp on the Euclidean plane.

Several variants of \tsp have been studied, most of which have a min-sum objective. One related problem with min-max
objective is {\em nurse station location}~\cite{Even2004}, where the goal is to obtain a collection of trees (each
rooted at a distinct depot) such that all vertices are covered and the maximum tree length is minimized. Even et
al.~\cite{Even2004} gave a 4-approximation algorithm for this problem. This is based on partitioning the \mst and
assigning to trees along the lines of Section~\ref{sec:previousTech}; their second step, however, involves a non-trivial bipartite matching subproblem.

In proving the properties of \levelPrim, we use {\em Light Approximate Shortest-Path Trees} introduced by Khuller, Raghavachari and Young~\cite{KRY95}, building on the work on shallow-light trees of Awerbuch, Baratz and Peleg~\cite{shallowlight}. An
$(\alpha,\beta)$-\last is a rooted tree that has (a) length at most $\beta$ times the \mst and (b) the distance from any
vertex to the root (along the tree) is at most $\alpha$ times the distance in the original metric. Khuller et
al.~\cite{KRY95} showed that every metric has an $\left(\alpha, 1+\frac{2}{\alpha-1}\right)$-\last (for any $\alpha>1$) and this is best possible.

One phase of our algorithm uses some ideas from scheduling on parallel machines~\cite{lenstraST}, which also has a
min-max objective. In this problem, job $j$ has processing time $p_{ij}$ on machine $i$ and the goal is to assign jobs to machines while minimizing the maximum completion time. Lenstra et al.~\cite{lenstraST} gave an LP-based
2-approximation algorithm for this problem.

\section{Model and Preliminaries} The input to the {\em Heterogenous TSP} (\htsp) consists of a metric $(V,d)$ denoting distances
between vertices, a depot $r\in V$ and $k$ vehicles with speeds $\{\lambda_i\}_{i=1}^k$ greater than or equal to 1. The goal is to find tours
$\{\tau_i\}_{i=1}^k$ (starting and ending at $r$) for each vehicle so that every vertex is covered in some tour and which
minimize the {\em maximum completion time} $\max_{i=1}^k\, \frac{d(\tau_i)}{\lambda_i}$.


At the loss of a factor of two in the approximation, we assume that the $\lambda_i$'s are all (non-negative integral) powers of $2$. Then, for each integer $i\ge 0$ we use $\mu_i$ to denote the number of vehicles with speed $2^i$. We let \opt
denote the optimal value of this modified instance of \htsp.

We let $G=(V,E)$ be the complete graph on vertices $V$ with edge-weights corresponding to the distance function $d$. For any set $F\sse E$ of edges, we set $d(F)=\sum_{e\in F}d_e$.
Given any (multi)graph $H$ and a subset $U$ of its vertices, $H[U]$ denotes the subgraph induced on $U$ and $H/U$
denotes the graph obtained by contracting vertices $U$ to a single vertex (we retain parallel edges). Moreover, for any pair of vertices $u,v$ in $H$, we use $d_H(u,v)$ to denote the length of the shortest path in $H$ between $u$ and $v$.

\ignore{ The setup of the problem is that we have $\mu_i$ trucks of speed $\lambda_i$ and we want to serve cities in
order to minimize the maximum load on the trucks.

In this problem we are given a set of points $P$ together with a special point $r$ called the root. We also have a
metric distance function $d(.,.)$ over $P \cup r$; in addition, we have values $\{\lambda_i\}$ and $\{\mu_i\}$. We want
to to construct trees $\{T_i^j\}_{j = 1}^{\mu_i}$ rooted at $r$ which cover all points in $P$. The cost of the tree
$T_i^j$ is $d(T_i^j) / \lambda_i$, where $d(T_i^j)$ is the sum of the distance (weight) of the edges in $T_i^j$, and
the cost of the cover is $\max_{i,j} cost(T_i^j)$. Our goal is to find a cover of minimum cost. }


\section{Algorithm for \htsp}

Assume that we have  correctly guessed a value $M$ such that $\frac{M}2\le \opt\le M$. (This value can be found via binary search and we address this in the end of Section \ref{sec:decomp}.)  We partition the set of vertices $V$ according to their distance to $r$:
$$V_0 = \{ u \in V : d(r, u) \le M\}, \mbox{ and}$$
$$V_i = \{u \in V : d(r, u) \in (2^{i - 1} M, \, 2^i M]\}, \mbox{ for all } i\ge 0.$$
%
The vertices in $V_i$ are referred to as {\em level $i$} vertices. For any $i\ge 0$, we use $V_{\le i}$ as a shorthand for $\cup_{j=0}^i V_j$ and similarly $V_{<i} = \cup_{j=0}^{i-1} V_j = V_{\le i-1}$.

We also
define the {\em level of an edge} $(u,v)\in E$ as the larger of the levels of $u$ and $v$. For each $i\ge 0$, $E_i$ denotes
the edges in $E$ of level $i$. Note that $d_e\le 2^{i+1} M$ for all $e\in E_i$, since both end-points of $e$ are
in $V_{\le i}$ and the triangle inequality bounds its distance by the two-hop path via the root. We use the notation $E_{\le i} = \cup_{j=0}^i E_j$ and $E_{\ge i}=\cup_{j\ge i} E_j$.

\subsection{Assignable Trees}

We start by studying collections of trees that can be assigned to vehicles in a way that each vehicle takes time $O(M)$ to visit all of its assigned trees.
%
%
%
\begin{definition}[Assignable Trees] \label{def:assignable}
A collection of $r$-rooted trees  $\bigcup_{i\ge 0} \T_i$ covering all vertices $V$ is called $(\alpha,\beta)$-\emph{assignable} if it
satisfies the following properties.
\begin{enumerate}
\initOneLiners
\item For each  $i\ge 0$ and every $T \in \T_i$,  $d(T) \le \alpha \; 2^i\, M$.
\item For each $i\ge 0$, $\sum_{j \ge i} d(\T_j) \le \beta M \sum_{j \ge i - 1} 2^j \, \mu_j$.
\end{enumerate}
\end{definition}

Intuitively, the trees in $\T_i$ can be assigned to vehicles with speed $2^i$ so as to complete in time $O(\alpha M)$.
Condition (2) guarantees that the trees $\bigcup_{j \ge i} \T_i$ targeted by vehicles of speed $2^i$ and above stand a chance of being handled by them within makespan $O(\beta M)$. Interestingly, these minimal conditions are enough to eventually assign all trees in collection to vehicles while guaranteeing makespan $O((\alpha + \beta) M)$.

\begin{lemma} \label{lem:assignable}
Given an assignable collection $\bigcup_{i\ge 0} \T_i$ of $r$-rooted trees, we can obtain in polynomial time an $(4\alpha + 2\beta)$-approximation for \htsp.
\end{lemma}

To prove this lemma\footnote{We remark that a direct proof of Lemma \ref{lem:assignable} is also possible, but the route we take reveals more properties of the requirement at hand and could potentially be useful in tackling generalizations of \htsp.}, we show that condition (2) guarantees the existence of a fractional assignment of trees where each vehicle incurs load at most $\beta M$. Then using condition (1) and a result on scheduling on parallel machines~\cite{lenstraST}, we round this assignment into an integral one while increasing the load on each vehicle by at most $2 \alpha M$. We loose an extra factor of 2 to convert the trees into routes.

\paragraph{Fractional Assignment.} Consider the bipartite graph $H$ whose left side contains one node for each tree in $\bigcup_i \T_i$ and whose right side contains one node for each vehicle. (We identify the nodes with their respective trees/vehicles.) There is an arc between the tree $T \in \T_i$ and a vehicle of speed $2^j$ if $j \ge i - 1$.

Consider the following $b$-matching problem in $H$: for each tree $T \in \T_i$, we set $b(T) = d(T)$ and for each vehicle $u$ of speed $2^j$ we set $b(u) = \beta 2^j M$. A (left-saturating) $b$-matching is one which fractionally assigns all $b(T)$ units of each tree $T$ such that no vehicle $u$ is assigned more than $b(u)$ units. Notice that a feasible $b$-matching gives a fractional assignment of trees where each vehicle incurs load at most $\beta M$.

Then our goal is to show the existence of a $b$-matching in $H$. Using a standard generalization of Hall's Theorem (e.g., see page 54 of \cite{4bills}), we see that $H$ has a feasible $b$-matching iff for every set $V$ of trees, $\sum_{T \in V} b(T)$ is at most $\sum_{u \in N(V)} b(u)$, where $N(V)$ is the neighborhood of $V$. However, the structure of $H$ allows us to focus only on sets $V$ which are equal to $\bigcup_{j \ge i} \T_j$ for some $i$.\footnote{To see that all other inequalities are dominated by those coming from such sets, first notice that if $V$ contains a tree in $\T_i$ then $N(V)$ already contains all vehicles of speed $2^j$ for $j \ge i - 1$. Then adding to $V$ extra trees in $\bigcup_{j \ge i} \T_j$ does not change its neighborhood and thus leads to a dominating inequality.} Using this revised condition, $H$ has a $b$-matching iff for all $i$, $\sum_{j \ge i} d(\T_j) \le \beta M \sum_{j \ge i - 1} 2^j \mu_j$. Since this is exactly condition (2) in Definition \ref{def:assignable}, it follows that $H$ indeed has a $b$-matching (which can be obtained in polynomial time using any maximum flow algorithm~\cite{4bills}).

\paragraph{Scheduling Parallel Machines.} We show how to round the fractional assignment obtained in the previous section. We consider each tree as a ``job'' and each vehicle as a ``machine", where the ``processing time'' $p_{T,u}$ of a tree $T$ in a vehicle $u$ of speed $2^j$ is $d(T)/2^j$; then the ``makespan'' of a vehicle is exactly equal to the sum of the processing times of the trees assigned to it.

Let $x_{T,u}$ denote the fraction of tree $T$ assigned to vehicle $u$ given by scaling down a $b$-matching in $H$ (i.e., if the matching assigns $d$ units of $T$ to vehicle $u$, we have $x_{T,u} = d/d(T)$). The feasibility of the matching gives $\sum_{T} x_{T,u} p_{T,u} \le \beta M$ for all $u$. Moreover, by construction of the edges of $H$, $x_{T,u} > 0$ for $T \in \T_i$ implies that $u$ has speed at least $2^{i - 1}$. Then using property (1) of assignable trees we get that $x_{T,u} > 0$ implies $p_{T,u} \le 2 \alpha M$. These two properties guarantee that $x$ is a feasible solution for the natural LP formulation for the scheduling problem with a feasible makespan value of $\beta M$ and the maximum processing time $t$ set to $2\alpha M$. Theorem 1 of \cite{lenstraST} then asserts that $x$ can be rounded into an integral assignment of trees to vehicles such that the load on any vehicle is at most $(2\alpha + \beta) M$.

As in Section \ref{sec:previousTech}, we can transform each tree in $\bigcup_{i \ge 0} \T_i$ into a cycle while at most doubling its length, which then gives a $(4 \alpha + 2 \beta)$ approximation for \htsp. This concludes the proof of Lemma \ref{lem:assignable}.


\subsection{\levelPrim} \label{sec:levelPrim}
\def\h{\mathcal{H}}

In order to obtain an assignable collection of $r$-rooted trees for our instance, we formally introduce \levelPrim trees. These are the trees obtained by the the following procedure. 

\begin{algorithm}[ht!]
  \label{alg:stagei}
  \caption{Level-Prim($G$)}
  \begin{algorithmic}[1]
    \STATE For each $i \ge 0$, let $H_i$ be an \mst for $G[V_{\le i}] / V_{< i}$.
    \RETURN $\h = \bigcup_{i \ge 0} H_i$.
    \end{algorithmic}
\end{algorithm}

Note that \levelPrim trees can alternately be defined by modifying Prim's algorithm such that nodes in level $i$ are only considered to be added to the tree after all nodes in levels below $i$ have already been added. 


\begin{theorem}
\label{thm:lPrim} A \levelPrim tree $\h=\{H_i\}_{i\ge 0}$ satisfies the following:
\begin{OneLiners}
 \item The vertex-levels along  every root-leaf path are non-decreasing.
 \item For each $i\ge 0$, $\sum_{j\ge i} d(H_j)\le 8\cdot \mst\left( G/ V_{< i} \right)$.
\end{OneLiners}
\end{theorem}


Note that the second property in Theorem~\ref{thm:lPrim} mirrors the second property in Definition~\ref{def:assignable}. A formal connection between the two is established via the following lemma that uses an optimal vehicle routing solution to derive a feasible spanning tree connecting a suffix of the level sets.

\begin{lemma}[Lower Bound] \label{lem:LB}
For each level $\ell\ge 0$, $\mst(G/V_{<\ell}) \le M\cdot \sum_{j\ge \ell-1} 2^j\, \mu_j$.
\end{lemma}
\begin{proof}
Consider an optimal solution for \htsp and let $E^*$ be the set of edges traversed by vehicles in this solution; label each edge in $E^*$ by the vehicle that traversed it. Clearly $E^*$ connects all vertices
to the root $r$.

Observe that only vehicles having speed at least $2^{\ell-1}$ can even reach any vertex in $V_{\ge \ell}$ (since a
vehicle of speed $s$ travels distance at most $s\cdot \opt\le s\cdot M$). Thus every edge in $E^*\cap E_{\ge \ell}$
must be labeled by some vehicle of speed at least $2^{\ell-1}$.
This implies that $d\left(E^*\cap E_{\ge
\ell}\right) \le M \cdot \sum_{j\ge\ell-1}2^j\, \mu_j$, since the right hand side is a bound on the total length traversed by vehicles having speed at least
$2^{\ell-1}$.

On the other hand, since $E^*$ connects all vertices, $E^*\cap E_{\ge \ell}$ contains a spanning tree of $G/V_{<\ell}$.
Thus we have $\mst(G/V_{<\ell}) \le d\left(E^*\cap E_{\ge \ell}\right) \le M\cdot \sum_{j\ge\ell-1}2^j\, \mu_j$.
\end{proof}

We then get the following corollary of Theorem~\ref{thm:lPrim}.

\begin{corollary}
\label{cor:lPrim} A \levelPrim tree $\h=\{H_i\}_{i\ge 0}$ satisfies the following:
\begin{OneLiners}
 \item The vertex-levels along  every root-leaf path are non-decreasing.
 \item For each $i\ge 0$, $\sum_{j\ge i} d(H_j)\le 8 M \sum_{j\ge i-1} 2^j\, \mu_j$.
\end{OneLiners}
\end{corollary}

In the rest of this section, we prove Theorem~\ref{thm:lPrim}. It is easy to see that for every $\ell$, $\bigcup_{j = 1}^\ell H_j$ spans $G[V_{\le \ell}]$, hence the procedure
produces a spanning tree for $G$. Moreover, by construction we obtain that every root-leaf path in $\h$ traverses the levels in non-decreasing order as desired. Thus, we focus on proving the second property in the theorem.

Instead of comparing the length of the edges in $\h$ with an $\mst$, it turns out to be much easier to use a specific \last tree as proxy for the latter. The following \last is implicit in the construction given in~\cite{KRY95}; for completeness we outline a proof in Appendix~\ref{app:spider}. Recall that a {\em spider} is a tree with at most one vertex (the center) having degree greater than two.

\begin{theorem}[\cite{KRY95}]\label{thm:last}
Given any metric $(V,d)$ with root $r$, there exists a spanning spider $L$ with center $r$ such that:
\begin{OneLiners}
\item For each $u\in V$, the distance from $r$ to $u$ in $L$ is at most $2\cdot d(r,u)$.
\item The length of $L$ is at most four times the \mst in $(V,d)$, i.e. $d(L)\le 4\cdot \mst$.
\end{OneLiners}
\end{theorem}

\begin{figure}[ht]
\centering
\subfigure[\mst on $G$]{
\includegraphics[scale=1]{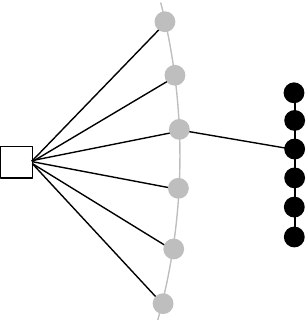}
\label{fig:subfig1}
}
\hspace{25pt}
\subfigure[A (1,2)-\last $T$ on $G$]{
\includegraphics[scale=1]{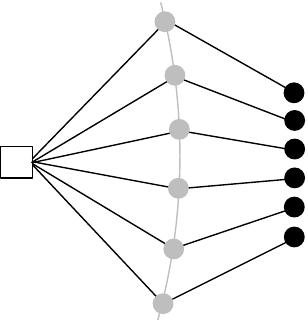} \hspace{15pt}
\label{fig:subfig2}
}

\caption{Instance with depot plus $n$ nodes in $V_0$ (box and gray nodes) and $n$ nodes in $V_1$ (black nodes). The distance between each distinct pair or nodes in $V_0$ is 1, the distance from a gray node to a black node is also 1 and the distance between two distinct black nodes is $1/n$. Notice that $d(T \cap E_1) = n$ while $\mst(G/V_0) < 2$.}
\label{fig:last}
\end{figure}

We remark that we cannot use a \last directly instead of \levelPrim since the former does not need to have the properties asserted by Theorem \ref{thm:lPrim}; it is easy to find a \last which does not satisfy the first property, while Figure \ref{fig:last} also shows that the second can also be violated by an arbitrary amount.
Using these spider {\last}s we can obtain the main lemma needed to complete the proof of Theorem \ref{thm:lPrim}.

\begin{lemma}\label{lem:lPrim-length}
For any graph $G$ and any \levelPrim tree $\h$ on $G$, we have $d(\h)\le 8\cdot \mst(G)$.
\end{lemma}
\begin{proof}
Consider a spider \last $L$ for $G$ and let \p denote the set of all root-leaf paths in $L$; note that $\p$ is edge-disjoint.

Consider any root-leaf path $P = (r = u_1 \rightarrow u_2 \rightarrow \ldots \rightarrow u_k)$ in \p. We claim that $P$
crosses levels almost in an increasing order. Specifically, there does not exist a pair of nodes $u_i, u_j \in P$ with
$i < j$, $u_i \in V_\ell$ and $u_j \in V_{\le \ell - 2}$. Suppose (for a contradiction) that this were the case; then
we would have that
$$d_L(r, u_j) = d_L(r, u_i) + d_L(u_i, u_j) \ge  d_L(r, u_i) \ge d(r,u_i) > 2^{\ell - 1} M,$$ where the last inequality uses
$u_i \in V_{\ell}$. On the other hand, $d(r,u_j)\le 2^{\ell-2}M$ since $u_j \in V_{\le \ell - 2}$; so we obtain
$d_L(r,u_j)>2\cdot d(r,u_j)$, which contradicts the definition of $L$ (see Theorem~\ref{thm:last}).

Now we transform $L$ into another spider $L'$ which traverses levels in non-decreasing order as follows. For each
root-leaf path $P = (r=u_1 \rightarrow u_2 \rightarrow \ldots \rightarrow u_k)$, perform the following modification.
Let $\{a_1, a_2, \ldots, a_{k'}\}$ be the subsequence of $P$ consisting of the vertices in {\em even numbered} levels,
i.e. each $a_i \in L_{2\ell}$ for some $\ell \ge 0$. Similarly, let $\{b_1, b_2, \ldots, b_{k''}\}$ be the subsequence
of $P$ consisting of the vertices in {\em odd numbered} levels. Define two paths  $P_{even} := (r \rightarrow a_1
\rightarrow \ldots \rightarrow a_{k'})$ (shortcutting $P$ over nodes $b_i$'s) and $P_{odd} = (r \rightarrow b_1
\rightarrow \ldots \rightarrow b_{k''})$ (shortcutting $P$ over $a_i$'s). Observe that both $P_{even}$ and $P_{odd}$
cross levels {\em monotonically}: if not then there must be some $i < j$ in $P$ with  $u_i \in V_\ell$ and $u_j \in
V_{\le \ell - 2}$, contrary to the previous claim. Also, by employing the triangle inequality we
have that $d(P_{even}), \,d(P_{odd}) \le d(P)$. Finally define the spider $L'$ as the union of the paths
$\{P_{even},\,P_{odd}\}$ over all root-leaf paths $P\in \p$.

By construction, the vertex levels along each root-leaf path of $L'$ are non-decreasing.  Additionally $d(L')
=\sum_{P\in \p} \left(d(P_{even})+d(P_{odd})\right) \le 2\, \sum_{P\in \p} d(P)=2 \cdot d(L) \le 8 \cdot \mst$, by
Theorem~\ref{thm:last}. Now partition the edges of $L'$ as:
$$\Delta_\ell = \left\{
\begin{array}{ll}
L'[V_{0}] & \mbox{ if }\ell=0,\\
L'[V_{\le \ell}]\setminus L'[V_{\le \ell-1}] & \mbox{ if }\ell\ge 1.
\end{array}
\right. $$

By the monotone property of paths in $L'$, it follows that $L'[V_{\le \ell}]$ is connected for every $\ell\ge 0$. Thus
$\Delta_\ell$ is a {\em spanning tree} in graph $G[V_{\le \ell}]/V_{<\ell}$. Since $H_\ell$ in the \levelPrim
construction, is chosen to be an \mst in $G[V_{\le \ell}]/V_{<\ell}$, we obtain $d(H_\ell)\le d(\Delta_\ell)$. So,
$d(\h) = \sum_{\ell\ge 0} d(H_\ell)\le \sum_{\ell\ge 0} \Delta_\ell = d(L')\le 8\cdot \mst$. This completes the proof
of the lemma.\end{proof}

\paragraph{Completing proof of Theorem~\ref{thm:lPrim}.} We now prove the second property in Theorem~\ref{thm:lPrim}.
Lemma~\ref{lem:lPrim-length} directly implies this property for $i=0$. For any level $i > 0$ consider the graph
$G'=G/V_{<i}$; observe that $\bigcup_{j\ge i} H_j$ is a \levelPrim tree for $G'$ (due to the iterative construction of
$\h=\bigcup_{\ell\ge 0} H_\ell$). Thus applying Lemma~\ref{lem:lPrim-length} to graph $G'$ and its \levelPrim
$\bigcup_{j\ge i} H_j$, we have $\sum_{j\ge i} d(H_j)\le 8\cdot \mst(G') = 8\cdot \mst(G/V_{<i})$.


\subsection{Decomposition Procedure}
\label{sec:decomp}

    In this section we decompose a \levelPrim tree into an assignable collection $\bigcup_{i \ge 0} \T_i$ of $r$-rooted trees. Motivated by Corollary \ref{cor:lPrim}, the idea is to essentially break each subgraph $H_i$ into many pieces and connect them to $r$ in order to form the set of trees $\T_i$. More specifically, assume for now that each connected component in $H_i$ is large enough, i.e. has length at least $2^i M$. Then for each $i \ge 0$, break the connected components of $H_i$ into trees of length approximately $2^i M$; add to each tree the shortest edge connecting them to $r$ and set $\T_i$ as the collection of $r$-rooted trees obtained. By construction we get that $\bigcup_{i \ge 0} \T_i$ satisfies the first property of an assignable collection. Moreover, notice that each edge added to connect a tree to the root has approximately the same length as the tree itself; this guarantees that $d(\T_i) \lesssim 2 d(H_i)$. It then follows that the collection $\bigcup_{i \ge 0} \T_i$ is assignable.

    Notice that it was crucial to break $H_i$ into trees of length at least approximately $2^i$. But this is problematic when $H_i$ has a small connected component. In this case
we show that such a small component is always attached to (``dangling'' from) a large enough component in $H_{i-1}$ (otherwise the dangling edge to a much earlier level will already make this component heavy enough not to be small); then we simply treat the small component as an integral part of the latter.

    Now we formally describe the proposed decomposition procedure.


{\bf Step 1.} Let $\subt_0$ contain the subtree $H_0=\h \bigcap E_0$. For each level $i\ge 1$: partition edges $\h
\bigcap E_i$ into a collection $\subt_i$ of (unrooted) subtrees such that each subtree contains exactly one edge from
$V_{<i}$ to $V_i$. For any $\tau \in \subt_i$ call the unique edge from $V_{<i}$ to $V_i$ its {\em head-edge}
$h(\tau)$. Note that such a partition is indeed possible since $\h[V_{\le i}]/V_{<i}$ is connected.

Any subtree in $\subt_i$ (for $i\ge 0$) is referred to as a level $i$ subtree. Note that head-edges are defined only
for subtrees in level $1$ and above.

{\bf Step 2.} For each level $i\ge 0$: {\em mark} those $\tau \in \subt_i$ that have $d(\tau)\ge 2^{i-3} M$. In addition, \emph{mark} the tree $H_0$ in $\subt_0$. Let $\subt^m_i$ and $\subt^u_i$ denote the marked and unmarked subtrees in $\subt_i$. 



{\bf Step 3.} For each level $i\ge 1$ and unmarked $\sigma \in \subt_i^u$: define $\pi(\sigma)$ as the subtree in
$\bigcup_{j<i} \subt_j$ containing the other end-point of $h(\sigma)$.
\begin{Myquote}
\begin{claim}\label{cl:decomp1} For $i\ge 1$ and unmarked $\sigma \in \subt_i^u$, $\pi(\sigma)\in\subt_{i-1}$. Moreover, $\pi(\sigma)$ is marked.
\end{claim}
\begin{proof}
Since $\sigma$ is unmarked in level $i\ge 1$, $d(h(\sigma))\le d(\sigma)< 2^{i-3} M$. So the end-point $v$ of
$h(\sigma)$ in $\pi(\sigma)$ satisfies $d(r,v)\ge \frac32\cdot 2^{i-2} M$, otherwise $d(h(\sigma)) \ge 2^{i-1}M-d(r,v)
> 2^{i-3}M$. In particular $v\in V_{i-1}$ and thus $\pi(\sigma)\in \subt_{i-1}$.

For the second part of the claim, notice that if $i = 1$ then $\pi(\sigma) = H_0$, which is always marked. So suppose $i\ge 2$. From the above, $\pi(\sigma)$ is in
level $i-1\ge 1$ and hence contains a head-edge. This implies that $\pi(\sigma)$ contains
some vertex $u\in V_{<i-1}$, namely an end-point of $h(\pi(\sigma))$. But then $d(\pi(\sigma))\ge d(u,v)\ge
d(r,v) - d(r,u)\ge 2^{i-3} M$, where we used $d(r,u)\le 2^{i-2}$ since $u\in V_{<i-1}$ and $d(r,v)\ge \frac32\cdot
2^{i-2} M$ from above. Thus $\pi(\sigma)$ must be marked.
\end{proof}
\end{Myquote}


{\bf Step 4.} For each level $i\ge 0$ and marked $\tau \in \subt_i^m$: define $\dangle(\tau)=\pi^{-1}(\tau)$ as the set
of all unmarked $\sigma\in \subt_{i+1}^u$ having $\pi(\sigma)=\tau$. Clearly $d(\sigma)\le 2^{i-2} M$ for all
$\sigma\in\dangle(\tau)$.

{\bf Step 5.} For each level $i\ge 0$ and marked $\tau \in \subt_i^m$: partition the tree $\tau \cup \dangle(\tau)$ into subtrees $T_1,\ldots,T_q$
such that the first $q-1$ trees have length in the range $[2^{i+1} M, 2^{i+2} M]$ and $T_q$ has length at most $2^{i+2} M$. Notice that this is possible since all edges of $\tau \cup \dangle(\tau)$ belong to $E_{\le i+1}$ and hence have length at most $2^{i + 1} M$. Finally, add the shortest edge from $r$ to each of these new subtrees to obtain a collection
$\T_i(\tau)$ of $r$-rooted trees.

\begin{Myquote}
\begin{claim}\label{cl:decomp2}
For any $T\in \T_i(\tau)$, we have $d(T)\le 3 \cdot 2^{i+1}M$.
\end{claim}
\begin{proof}
Notice that every $T \in \T_i(\tau)$ consists of a $T_j$ (for some $1\le j\le q$) and an edge from $r$ to a node in $V_{\le i + 1}$. Since the former has length at most $2^{i + 2} M$ and the latter has length at most $2^{i+1} M$, it follows that $d(T) \le 3 \cdot 2^{i+1} M$.
\end{proof}

\begin{claim} $\sum_{T\in \T_i(\tau)} d(T) \le 5 \cdot \left[d(\tau) + d(\dangle(\tau))\right]$. \label{cl:decomp3}
\end{claim}
\begin{proof}
We break the analysis into two cases depending of $q$. Suppose $q = 1$, namely $\T_i(\tau)$ consists of a single tree
$T$. In this case $T = \tau \cup \dangle(\tau) \cup \{e\}$, where $e$ is an edge to $r$. If $i = 0$ then $d(e) = 0$ and the result holds directly. If $i > 0$ then $\tau$ has a node in
$V_{< i}$ and hence $d(e) \le 2^{i - 1} M$. Because $\tau$ is
marked and different than $H_0$, the lower bound on its length implies that $d(e) \le 2^{i - 1} M\le 4 d(\tau)\le 4 (d(\tau) +
d(\dangle(\tau))$. The result follows by adding the length of $\tau \cup \dangle(\tau)$ to both sides.

Now suppose $q > 1$. Since all trees in $\T_i(\tau)$ lie in $V_{\le i +1}$, each edge from the root in $\T_i(\tau)$ has
length at most $2^{i+1}M$. So the left hand side is at most $\sum_{j = 1}^q d(T_j) + q \cdot 2^{i + 1} M$. But
for $j < q$ we have $d(T_j) \ge 2^{i + 1} M$, so the last term of the previous expression can be upper bounded
by $\frac{q}{(q-1)} \sum_{j=1}^{q-1} d(T_j)$. This bound is smallest when $q = 2$, which then gives $\sum_{T \in
\T_i(\tau)} d(T) \le 3 \sum_{j = 1}^q d(T_j) \le 3 d(\tau)$. This concludes the proof of the claim.
\end{proof}
\end{Myquote}

{\bf Step 6.} For each level $i\ge 0$: define $\T_i = \bigcup_{\tau\in \subt^m_i} \, \T_i(\tau)$.

\medskip
The following lemma summarizes the main property of our decomposition procedure.
\begin{lemma}\label{lem:isAssignable}
The collection $\{\T_i\}_{i\ge 0}$ obtained from the above procedure is $(6,40)$-assignable.
\end{lemma}

\begin{proof} By Claim~\ref{cl:decomp2}, each tree in $\T_i$ has length at most $3 \cdot 2^{i+1} M$. So the collection satisfies condition~(1) of Definition~\ref{def:assignable}.

Fix any $i\ge 0$ for condition (2) in Definition~\ref{def:assignable}. Due to Corollary \ref{cor:lPrim}, it suffices to prove that $\sum_{j\ge i} d(\T_j) \le 5\cdot \sum_{j\ge i}d(H_j)$. Using Claim~\ref{cl:decomp3} we
obtain that
$$d(\T_j) = \sum_{\tau\in \subt^m_j} d(\T_i(\tau)) \le 5 \cdot \sum_{\tau\in \subt^m_j} \left[ d(\tau) +
d(\dangle(\tau))\right] = 5\cdot d(\subt^m_j) + 5\cdot d(\subt^u_{j+1}).$$
The last equality above uses the fact that that $\{\dangle(\tau) : \tau\in \subt^m_j\}$ is a partition of
$\subt^u_{j+1}$. Thus:
\begin{equation}\label{eq:decomp-suffix}\sum_{j\ge i} d(\T_j) \le 5\cdot \sum_{j\ge i} d(\subt^m_j) + 5\cdot \sum_{j\ge
i} d(\subt^u_{j+1}) \le 5\cdot \sum_{j\ge i} d(\subt_{j}) = 5\cdot \sum_{j\ge i}d(H_j).\end{equation}
This concludes the proof of Lemma \ref{lem:isAssignable}.
\end{proof}

\paragraph{Summary of the Algorithm.} Our algorithm starts with an initial low guess of $M$ and runs the \levelPrim procedure. If the second condition in Corollary~\ref{cor:lPrim} does not hold for this run, we double the guess for $M$ and repeat until it is satisfied (this happens the first time that $M$ reaches the condition for the correct guess: $\frac{M}2\le \opt\le M$).
We use the decomposition in this section summarized in Lemma~\ref{lem:isAssignable} to obtain a (6,40)-assignable collection of trees. Using Lemma~\ref{lem:assignable} on this collection gives us the desired constant approximation ratio by observing that the guess $M$ in this step obeys $M \le 2 \cdot \opt$.


\section{Generalization for \hvrp} \label{sec:hvrp}

\newcommand{\is}{\ensuremath{\mathcal{I}}\xspace}
\newcommand{\js}{\ensuremath{\mathcal{J}}\xspace}
\newcommand{\optt}{\ensuremath{{\sf OPT_{tsp}}}\xspace}
\newcommand{\optv}{\ensuremath{{\sf OPT_{vrp}}}\xspace}
\def\st{{\ensuremath{\sf MinSt}}}

The input to the {\em Heterogenous CVRP} (\hvrp) consists of a metric $(V,d)$ denoting distances between vertices,
depot $r\in V$ (containing an infinite supply of items), demands $\{q_v\}_{v\in V}$ and $k$ vehicles with speeds
$\{\lambda_i\}_{i=1}^k$, each having capacity $Q$. A solution to \hvrp consists of tours $\{\sigma_i\}_{i=1}^k$
(starting and ending at $r$) for each vehicle so that all demands are satisfied and each vehicle carries at most $Q$
items at any point in time. The objective is to minimize the {\em maximum completion time}, $\max_{i=1}^k\,
\frac{d(\sigma_i)}{\lambda_i}$. We study the ``split-delivery'' version of \cvrp here, where demand at a vertex may be
served by multiple visits to it; however, our result easily extends to the ``unsplit-delivery'' \hvrp.

We show that the \hvrp problem can be reduced to \htsp in an approximation preserving way; so we also obtain an
$O(1)$-approximation for \hvrp. The idea in this reduction is to modify the input metric based on lower-bounds for
\cvrp~\cite{haimovichRK}. In order to avoid ambiguity, we use $\optv$ to denote the optimum for $\hvrp$ and $\optt$ to
denote the optimum for $\htsp$.

\begin{theorem}
\label{th:vrptotsp}
Consider an instance \is of \hvrp. There is a poly-time constructible instance \js of \htsp such that $\optt(\js) = O(1) \cdot \optv(\is)$. Moreover, a solution to \js of makespan $M$ can be converted in poly-time to a solution to \is with makespan $O(M)$. 
\end{theorem}
\begin{proof}
Let \is be an instance of \hvrp as specified above. Standard scaling arguments can be used to ensure that $Q$ is
polynomial in $n$ and $q_v\in\{0,1,\ldots,Q\}$ for all $v\in V$ (details in the full version).

Let $G=(V,E)$ denote the complete graph on vertices $V$ with edge-weights equal to distances $d$. Augment $G$ to a new
graph $H$ by adding vertices $V'=\{v_p : v\in V,\, p\in[q_v]\}$, and edges $E'=\{(v,v_p) : v\in V,\, p\in[q_v]\}$; each
edge $(v,v_p)$ has weight $\frac{d(r,v)}{Q}$. For any vertex $v\in V$, the vertices $\{v_p : p\in[q_v]\}$ are referred
to as copies of $v$.
Let $(V',\ell)$ denote the metric induced on vertices $V'$ where $\ell$ denotes the shortest-path distances in graph
$H$. We let \js be the instance of \htsp on metric $(V',\ell)$ with depot $r$ and $k$ vehicles having speeds
$\{\lambda_i\}_{i=1}^k$. Since $Q\le poly(n)$ this reduction runs in polynomial time.

For any graph $L$ and subset $S$ of vertices, let $\st_L(S)$ denote the minimum length Steiner tree connecting $S$. For
any subset $T\sse V'$ and $v\in V$ let $N_v(T)$ denote the number of $v$-copies in $T$; also define $\pi(T) = \{v\in V : N_v(T)>0\}$. Observe that for any $T\sse V'$ we have $\st_H(T)=\st_G(\pi(T))+\sum_{v\in V}
\frac{N_v(T)\cdot d(r,v)}{Q}$ by the definition of graph $H$.

We first show that $\optt(\js) = O(\optv(\is))$. Consider an optimal solution $\{\sigma_i\}_{i=1}^k$ to \is. For each
$i\in[k]$, let $c_i(v) \in [q_v]$ denote the units of demand at vertex $v\in V$ served by vehicle $i$, and let $S_i =
\{v\in V : c_i(v)>0\}$. Note that $\sum_{i=1}^k c_i(v)=q_v$ for all $v\in V$; hence we can choose $S'_i\sse V'$ for
each $i\in [k]$ such that $\cup_{i=1}^k S'_i = V'$ and $N_v(S'_i)=c_i(v)$ for all $v\in V,\, i\in [k]$. Since
$\sigma_i$ is a capacitated tour in $G$ serving demands $\{c_i(v) : v\in S_i\}$, we have $d(\sigma_i)\ge
\max\left\{\st_G(\{r\}\cup S_i),\, \sum_{v\in S_i} \frac{c_i(v)\cdot d(r,v)}{Q}\right\}$ using the (connectivity and
capacitated routing) lower-bounds for \cvrp~\cite{haimovichRK}. Thus $\st_H(\{r\}\cup S'_i) = \st_G(\{r\}\cup S_i) +
\sum_{v\in S_i} \frac{c_i(v)\cdot d(r,v)}{Q} \le 2\cdot d(\sigma_i)$. Now consider the solution to \js where the
$i^{th}$ vehicle visits vertices $S'_i$ along the minimum TSP tour on $\{r\}\cup S'_i$, for all $i\in [k]$; the
distance traversed by the $i^{th}$ vehicle is at most $2\cdot \st_H(\{r\}\cup S'_i)\le 4\cdot d(\sigma_i)$. So the
\htsp objective value of this solution is at most $\max_{i\in [k]} \frac{4\cdot d(\sigma_i)}{\lambda_i}=4\cdot
\optv(\is)$.

Now consider a solution $\{\tau_i\}_{i=1}^k$ to \js with makespan $M$. Let $R_i\sse V'$
denote the vertices that are served by each vehicle $i\in[k]$. Since $\tau_i$ is a TSP
tour on $\{r\}\cup R_i$, we have  $d(\tau_i)\ge \st_H(\{r\}\cup R_i) = \st_G(\{r\}\cup \pi(R_i)) + \sum_{v\in V}
\frac{N_v(R_i)\cdot d(r,v)}{Q}$. Now fix $i \in [k]$ and consider the instance of \cvrp on vertices $\{r\}\cup
\pi(R_i)$ with demands $\{N_v(R_i): v\in \pi(R_i)\}$. As mentioned in the previous paragraph, $\max\left\{\st_G(\{r\}\cup \pi(R_i)),\, \sum_{v\in
\pi(R_i)} \frac{N_v(R_i)\cdot d(r,v)}{Q}\right\}$ is a lower-bound for this instance, and the algorithm from~\cite{haimovichRK} returns a solution $\sigma_i$ within a $\rho=O(1)$ factor of this lower-bound. It readily follows that $\{\sigma\}_{i = 1}^k$ is a feasible solution to \is with makespan at most $\max_{i\in [k]} \frac{\rho\cdot d(\tau_i)}{\lambda_i}= O(M)$.
\end{proof}

We note that this algorithm returns a {\em non-preemptive} \hvrp solution, i.e., each item once picked up at the depot
stays in its vehicle until delivered to its destination. Moreover, the lower-bounds used by the \hvrp algorithm also
hold for the (less restrictive) {\em preemptive} version, where items might be left temporarily at different vertices
while being moved from the depot to their final destination. Thus our algorithm also bounds the ``preemption gap''
(ratio of optimal non-preemptive to preemptive solutions) in \hvrp by a constant.


\section{Open Problems}


One interesting open question regards the approximability of $\htsp$ and $\hvrp$ when vehicles are located in multiple
different depots across the space. The current definition of an assignable collection and the definition of \levelPrim
crucially depend on the assumption of a unique depot, hence an extension to the multi-depot case is likely to require
new ideas. Another interesting direction is to consider \hvrp with non-uniform capacities, where the ideas presented in
Section \ref{sec:hvrp} do not seem to generalize directly.


\pagebreak

\bibliographystyle{alpha}
{\small \bibliography{hetVRP}}

\pagebreak

\appendix

\section{Proof of  Theorem \ref{thm:last}} \label{app:spider}


We will show that the following algorithm produces an $(\alpha, \beta)$-spider for $G$.

\begin{algorithm}[ht!]
  \label{alg:spider}
  \caption{$(\alpha,\beta)$-Spider}
  \begin{algorithmic}[1]
    \STATE Consider an \mst for $G$ and traverse it in preoder to obtain a path $S^0 = S = (r = u_1 \rightarrow u_2 \rightarrow \ldots \rightarrow u_n)$ such that $d(S^0) \le 2 \cdot \mst$.
    \FOR{$i$ from 1 to n}
        \IF{$d_S(u_1, u_i) > \alpha d(u_1, u_i)$} \label{alg:shortcutIf}
            \STATE {\bf add} $(u_1, u_i)$ to $S$ and {\bf mark} $u_i$. \label{alg:shortcut}
        \ENDIF
    \ENDFOR
    \STATE {\bf for each} marked node $u_i$, remove $(u_{i -1}, u_i)$ from $S$.
    \label{alg:delete}
    \RETURN $S^* = S$.
    \end{algorithmic}
\end{algorithm}

\begin{lemma}
 The graph $S^*$ returned by the algorithm is a spider.
\end{lemma}

\begin{proof}
    Since the algorithm keeps adding edges to the path $S^0$, it is clear that before Step~\ref{alg:delete} only the root $u_1$ and marked nodes have degree larger than 2. Moreover, each marked node has degree exactly 3. Thus, after Step~\ref{alg:delete} we have that only the root has degree larger then 2, and the lemma follows.
\end{proof}

%

\begin{lemma}
    $S^*$ is an $(\alpha, \beta)$-spider.
\end{lemma}

\begin{proof}
    First we prove that $d_{S^*}(u_1, u_i) \le \alpha d(u_1, u_i)$ for all $i$. To see this, consider $S$ right before Step \ref{alg:delete}. It follows from Step \ref{alg:shortcut} that $d_S(u_1, u_i) \le \alpha d(u_1, u_i)$ for all $i$. Noticing that $S^*$ is a shortest path tree of $S$ from node $u_1$ implies the desired result.

    Now we prove that $S^*$ satisfies the second property of $(\alpha, \beta)$-spider. Define $v_0 = u_1$ and let $v_i$ be the $i$th node marked by the algorithm. It is clear that $d(S^*) \le d(S^0) + \sum_{i = 1}^k d(u_1, v_i)$; so our goal is to upper bound the last summation.

    Fix a node $v_i$. Consider the beginning of the iteration where $v_i$ is marked. Notice that at this point $d_S(u_1, v_i) \le d(u_1, v_{i-1}) + d_S(v_{i-1}, v_i)$, since edge $(u_1, v_{i-1})$ was already added to $S$; since $S^0$ is subgraph of $S$, it is also clear that the right hand side is at most $d(u_1, v_{i-1}) + d_{S^0}(v_{i-1}, v_i)$. However, since $v_i$ was marked, we have that $d_S(u_1, v_i) > \alpha \cdot d(u_1, v_i)$, and then using the previous bounds we obtain that $\alpha \cdot d(u_1, v_i) < d(u_1, v_{i-1}) + d_{S^0}(v_{i-1}, v_i)$.

    Adding the previous inequality over all $v_i$'s we get that $\alpha \sum_i d(u_1, v_i) < \sum_i d(u_1, v_{i-1}) + \sum_i d_{S^0}(v_{i-1}, v_i).$ Noticing that $d(u_1,v_0) = d(u_1, u_1) = 0$ and reorganizing leads to $(\alpha - 1) \sum_i d(u_1, v_i) \le \sum_i d_{S^0}(v_{i-1}, v_i)$. Finally, notice that $\sum_{i = 1}^k d_{S^0}(v_{i-1}, v_i) \le d(S^0)$: this follows from traversing the path $S^0$ and using the triangle inequality. This gives the bound $\sum_i d(u_1, v_i) \le d(S^0)/(1-\alpha)$.

    Plugging this back to a previous bound on the length of $S^*$ gives $d(S^*) \le (1 + 1/(1 - \alpha))d(S^0) \le (2 + 2/(1- \alpha)) d(S^0)$. This concludes the proof of  Theorem \ref{thm:last}.
\end{proof}

\end{document}